\documentclass{mem}
\usepackage{natbib}\usepackage{txfonts}\usepackage{balance}
\usepackage{graphicx}
\usepackage[a4paper]{hyperref}
\idline{75}{282}
\begin{document}
\def\teff{$T\rm_{eff }$}
\def\kms{$\mathrm {km s}^{-1}$}

\title{
Search for SiO masers in nearby Miras pulsating in the first overtone mode
}

   \subtitle{}

\author{
Yoshifusa \,Ita\inst{1}, Shuji\, Deguchi\inst{2}, Noriyuki\, Matsunaga\inst{3} \and Hinako\, Fukushi\inst{3}
          }

  \offprints{Y. Ita}

\institute{
Institute of Space and Astronautical Science, Japan Aerospace Exploration Agency, 3-1-1 Yoshinodai, Sagamihara, Kanagawa 229-8510, Japan \\
\email{yita@ir.isas.jaxa.jp}
\and
Nobeyama Radio Observatory, National Astronomical Observatory, Minamimaki, Minamisaku, Nagano 384-1305, Japan
\and
Institute of Astronomy, School of Science, The University of Tokyo, 2-21-1 Osawa, Mitaka, Tokyo 181-0015, Japan
}

\authorrunning{Y. Ita}

\titlerunning{SiO Masers in nearby Miras pulsating in the 1st overtone mode}

\abstract{
We studied the period-K magnitude (P$-$K) relations of nearby Mira and Mira-like variables with relatively good Hipparcos parallaxes. They form at least two prominent sequences on the P$-$K plane, corresponding to the sequences C (Mira variables pulsating in the fundamental mode) and C$^\prime$ (Mira variables pulsating in the 1st overtone mode), that were found in the LMC. As a part of an ongoing study to see the differences between the Mira variables pulsating in the fundamental and the 1st overtone mode, we searched for SiO masers in the nearby variables on the sequences C$^\prime$ and C using the Nobeyama 45m radio telescope. We observed 28 selected nearby Mira and Mira-like variables without previous maser observations, and found 3 new emitters. The observational result shows that there is few or no SiO maser emitters pulsating in the 1st overtone mode.

}
\maketitle{}

\section{Introduction}
After the pioneering work by \citet{wood2000}, variable stars in the Large and Small Magellanic Clouds (LMC and SMC) have been studied by many authors (e.g., \citealt{cioni2001}, \citealt{noda2003}, \citealt{kiss2003}, \citealt{ita2004a}, \citealt{gro2004}, \citealt{sos2004}, \citealt{raimondo2005}, \citealt{fraster2005}), with each author using different time-series photometric database (i.e., MACHO, OGLE, MOA, EROS).

\citet{ita2004a} suggested that there are at least two dominant pulsation modes among Mira variables. One is the fundamental mode, and the other is the 1st overtone mode. Then, what determines the dominant pulsation mode of Mira variables? Comparing various observable quantities between the two groups will give us clues to answer the question. Following this idea, \citet{ita2004b} compared the NIR photometric properties between the two modes and showed that: (1) they separate on the period-$J - K$ colour diagram; (2) the pulsation amplitudes of the 1st overtone Miras are smaller than those of the fundamental ones.

To know their differences in more detail, the imaging data alone is no longer sufficient. The spectroscopic data, radio data,  mid-IR data and so on would bring us new insights. In this regard, the Magellanic Clouds (MCs) are technically a bit far for such observations. Therefore, we decided to leave the MCs and revisit our own Galaxy. Throughout this paper, we will use the word ``Mira-like variables'' instead of ``Semi-Regular variables'', because the boundary is blurred between the Miras and SRa variables.

\section{Observation and Result}
\subsection{Source Selection}
Because the Galactic variables are near, we can observe them in every possible wavelengths and with any desired observational styles. However, there is a critical disadvantage when compared with studying variable stars in the MCs; the distances to the Galactic objects are usually difficult to know. Therefore, we restricted our sample stars to the Mira and Mira-like variables with relatively good Hipparcos parallaxes \citep{esa1997}.

The target stars are selected from \citet{bedding1998}, \citet{whitelock2000} and \citet{knapp2003} on conditions that: (1) The spectral types are M or S; (2) $\frac{\Delta\omega}{\omega} <$ 0.44, where $\omega$ and $\Delta\omega$ are the parallax and its error, respectively. Ninety five stars satisfy these criteria. Among them, 28 stars are observable from Nobeyama ($\delta$ $>$ 30$^\circ$) and without previous SiO maser observations.

\subsection{Observation using Nobeyama 45m radio telescope and the Result}
As a part of an ongoing study to see the intrinsic differences between the fundamental and the 1st overtone Mira, we searched for SiO masers (\textit{J}=1-0, \textit{v}=1 and 2 at the rest frequencies of 43.122 and 42.821 GHz, respectively) towards the above 28 selected nearby Mira and Mira-like variables. The observation was carried out with Nobeyama 45m radio telescope in June 2005. As a result, we found 3 new SiO maser emitters (AS Her, BG Cyg and UU Dra). Details of the observation will be described in the future paper (Ita et al. in preparation).

\begin{figure}
\includegraphics[angle=-90,scale=0.26]{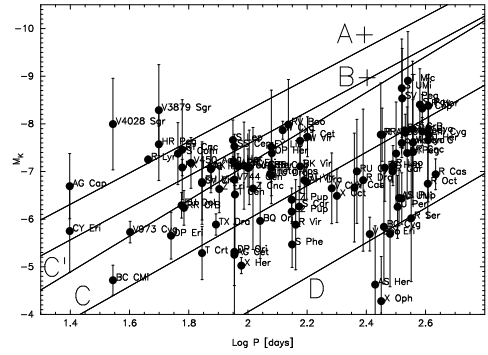}
\caption{\footnotesize P$-$K diagram of nearby oxygen-rich Mira variables with M/S spectral type and relatively good ($\frac{\Delta\omega}{\omega} <$ 0.44) Hipparcos parallaxes. The lines show the sequences of variable stars in the LMC as given by \citealt{ita2004a}, whose zero-points are shifted by 18.5 mag to account for the distance to the LMC.}
\label{fig1}
\end{figure}

\section{Discussion}
\subsection{P$-$K diagram of nearby Miras}
Fig.\ref{fig1} is the P$-$K diagram of nearby oxygen-rich Mira and Mira-like variables with relatively good Hipparcos parallaxes. Note that this figure includes stars that cannot be observed from Nobeyama. The pulsation periods and K magnitudes were taken from \citet{bedding1998}, \citet{whitelock2000} and \citet{knapp2003}. For the star SV Peg, the General Catalog of Variable Stars (GCVS, \citealt{kholopov1992}) lists its pulsation period of 144.6 days, but we adopt 332.0 days by analyzing data from the VSOLJ\footnote{http://www.kusastro.kyoto-u.ac.jp/vsnet/VSOLJ/vsolj.html}. Due to the errors in Hipparcos parallaxes and the large dispersions in mass and metallicity, we don't see the tight P$-$K relations as those in the LMC. However, it is clear that the Mira and Mira-like variables in our Galaxy are likely to pulsate in the wide variety of pulsation modes just like the LMC variables. Also, there seems to be many Galactic counterparts of Mira variables pulsating in the 1st overtone mode.

\begin{figure*}[t!]
\begin{center}
\includegraphics[angle=-90,scale=0.43]{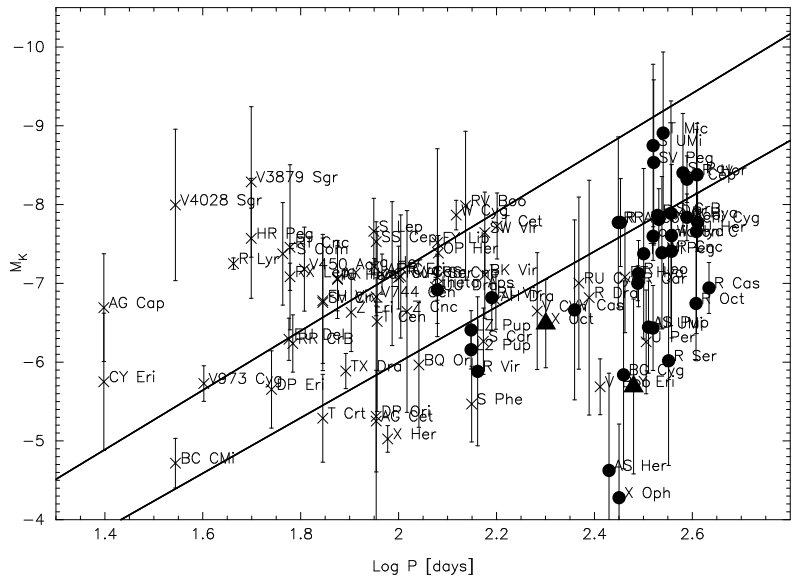}
\caption{\footnotesize The same as the Fig.\ref{fig1}, but with the SiO maser detection status. Filled circles represent detections, crosses non-detections and filled triangles southern stars (cannot be observed from Nobeyama, and no information is available in literature).}
\label{fig2}
\end{center}
\end{figure*}

\begin{figure}
\includegraphics[angle=-90,scale=0.26]{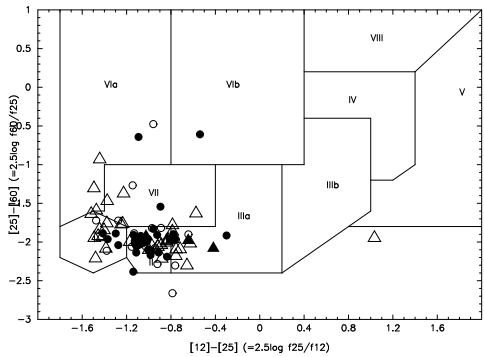}
\caption{\footnotesize IRAS two color diagram of sample stars. See \citet{veen88} for the definition of the background regions. The circles show fundamental mode pulsators and triangles 1st overtone mode ones. The filled/open symbols represents SiO maser detections/non-detections.}
\label{fig3}
\end{figure}

\subsection{SiO maser emission and the pulsation modes}
\citet{patel1992} studied the relationship between the SiO maser emission and the intrinsic properties of the Mira variables. They concluded that there seem to be two criteria which may inhibit a Mira variable from giving rise to SiO maser emission: (1) Spectral type earlier than M6; (2) Bolometric magnitude fainter than $-$4.8 mag. Also, it has been known from our SiO maser survey project (e.g., \citealt{deguchi2004}) that the maser detection rate goes up with increasing pulsation period.

After these studies, we added the results of SiO maser observations to the Fig.\ref{fig1} to see whether the SiO maser detection/non-detection depends on the pulsation modes. Fig.\ref{fig2} shows the result. Note that this figure includes stars that cannot be observed from Nobeyama, but whose SiO maser (43 \& 86 GHz) detection/non-detection can be known in the literature (\citealt{benson1990}, \citealt{alcolea1990}, \citealt{haikala1994}, \citealt{jiang1996}, \citealt{gonzalez2003}). It can be seen that the SiO masers are mainly detected from the fundamental mode pulsators. This is an apparently conflicting result with \citet{patel1992}, who suggested that a large value of radius may allow a larger path length for the gain of the maser, and hence a greater maser luminosity. However, our observational result does not favor this idea, because the 1st overtone mode pulsators should have larger radii than those of the fundamental mode ones, if compared with the same pulsation period (e.g., figure 1 in \citealt{whitelock2000}).

To determine the dominant pulsation mode of each star more quantitatively, we calculated the direct distance
 from each data point (Log P, K) to the sequences C and C$^{\prime}$, and adopted the nearer corresponding pulsation mode. The Fig.\ref{fig3} is the IRAS two color diagram of sample stars showing the correlation between the SiO maser emission and the mass-loss and also the pulsation modes. It is believed that SiO maser emission is strongly connected to the mass-loss. However, we don't see such strong correlation in the figure. Instead, the SiO maser emission is likely to be more of a pulsation mode issue.

Unfortunately, there are few or almost no long-period 1st overtone Miras in our sample (in fact, there are few or almost no long-period 1st overtone Miras within the Hipparcos's range). Therefore, we cannot conclude that the fundamental pulsation mode is the necessary condition for SiO maser emission. We have to wait for the results of GAIA satellite till when she will drastically increase the number of galactic Mira variables with good parallaxes.

\section{Conclusions}
The relevance between the SiO maser emission and the pulsation modes of nearby Mira variables were studied by adding the new observational data (3 detections and 25 non-detections) to the previous studies. As far as the present data suggests, it is likely that the SiO maser emission favors the fundamental pulsation mode.


\begin{acknowledgements}
This paper makes use of the data from the Hipparcos satellite, IRAS satellite, GCVS and the VSOLJ database. This research is supported in part by the Grant-in-Aid for Encouragement of Young Scientists (B) No. 17740120 from the Ministry of Education, Culture, Sports, Science and Technology of Japan.
\end{acknowledgements}

\bibliographystyle{aa}

\end{document}